\documentclass[aps,prl,twocolumn,superscriptaddress,showpacs]{revtex4}

\usepackage{mathrsfs}
\usepackage{graphicx}
\usepackage{color}
\usepackage{amsmath}
\usepackage{amssymb}

\begin{document}

\title{Localization and universal fluctuations in ultraslow diffusion processes}

\author{Alja\v{z} Godec}
\affiliation{Institute for Physics \& Astronomy, University of Potsdam,
14476 Potsdam-Golm, Germany}
\affiliation{National Institute of Chemistry, 1000 Ljubljana, Slovenia}
\author{Aleksei V. Chechkin}
\affiliation{Institute for Theoretical Physics, Kharkov Institute of Physics
and Technology, Kharkov 61108, Ukraine}
\affiliation{Max-Planck Institute for the Physics of Complex Systems,
N{\"o}thnitzer Stra{\ss}e 38, 01187 Dresden, Germany}
\affiliation{Institute for Physics \& Astronomy, University of Potsdam,
14476 Potsdam-Golm, Germany}
\author{Eli Barkai}
\affiliation{Department of Physics, Bar Ilan University, Ramat-Gan 52900,
Israel}
\author{Holger Kantz}
\affiliation{Max-Planck Institute for the Physics of Complex Systems,
N{\"o}thnitzer Stra{\ss}e 38, 01187 Dresden, Germany}
\author{Ralf Metzler}
\affiliation{Institute for Physics \& Astronomy, University of Potsdam, 14476
Potsdam-Golm, Germany}
\affiliation{Department of Physics, Tampere University of Technology, FI-33101
Tampere, Finland}


\begin{abstract}
We study ultraslow diffusion processes with logarithmic mean squared displacement
(MSD) $\langle x^2(t)\rangle\simeq\log^{\gamma}t$. Comparison of annealed
continuous time random walks (CTRWs) with logarithmic waiting time distribution
$\psi(\tau)\simeq1/(\tau\log^{1+\gamma}\tau)$ and Sinai diffusion in quenched
random landscapes shows striking similarities, despite their very different
physical nature. In particular, they exhibit a weakly
non-ergodic disparity of the time and ensemble averaged MSDs. Remarkably, for the
CTRW we observe that the fluctuations of time averages become universal with an
exponential suppression of mobile trajectories. We discuss the fundamental
connection between the Golosov localization effect and non-ergodicity in the
sense of the disparity between ensemble and time averaged MSD.
\end{abstract}


\maketitle

Ever since Karl Pearson's defining letter to the editor 1905 \cite{pearson} as
well as Einstein's and Smoluchowski's mean free path studies \cite{smolu},
\emph{random walks\/} have become a standard approach to a multitude of
nonequilibrium phenomena across disciplines \cite{hughes,sornette,gallager}.
Most frequently renewal random walks are used \cite{gallager} in which jumps
are independent of previous jumps, reflecting the motion in annealed environment
\cite{bouchaud}. These contrast random walks in quenched environments in which
a particle progressively builds up correlations when it returns to previously
visited locations with site-specific properties \cite{bouchaud}. The prototype
approach is Temkin's lattice model with site-dependent probabilities for jumping
left or right \cite{hughes,temkin}.

A great leap forward came with Sinai's study of a special case of Temkin's model
in which a walker jumps from site $x$ to $x\pm1$ with the site-specific probability
$p_x=\frac{1}{2}(1\pm\varepsilon s_x)$ \cite{sinai}. Here, the amplitude $0<
\varepsilon<1$ and $s_x=\pm1$ with probability $1/2$ \cite{hughes,gleb}. Sinai
diffusion can be viewed as a random walk in the quenched potential landscape
created by a standard random walk. A simple argument for the temporal spreading
in Sinai diffusion goes as follows \cite{bouchaud}. To span a distance $x$ from
its starting point the particle needs to cross an energy barrier of
order $\sqrt{x}$ with activation time $\tau\sim\tau_1\exp(\sigma\sqrt{x})$, where
$\sigma$ is is a measure for the disorder strength versus thermal energy and $\tau_1$ a fundamental
time scale. The
distance covered by the walker during time $t$ then scales as $x^2\simeq\ln^4(t/
\tau_1)$.

Sinai diffusion is related to the random-field Ising model \cite{ising,ising1}
and helix-coil phase boundaries in random heteropolymers \cite{helixcoil}. In a
biophysical context, due to the inherently quenched heterogeneity of
biomolecules Sinai-type models are used to describe mechanical DNA
unzipping \cite{unzipping}, translocation of biomolecules through nanopores
\cite{translocation,kafri}, and molecular motor motion \cite{kafri}. Ultraslow
diffusion with mean squared displacement (MSD)
\begin{equation}
\label{msd}
\langle x^2(t)\rangle\simeq2K_{\gamma}\ln^{\gamma}t, \,\,\, \gamma>0
\end{equation}
in fact has a much broader scope in highly disordered low-dimensional systems:
inter alia, in vacancy induced motion \cite{hill,olivier},
biased motion in exclusion processes \cite{juhasz},
motion in aging
environments \cite{lomholt}, compactification of paper \cite{paper} or grain \cite{grain},
dynamics in glassy systems \cite{glasses}, record statistics \cite{record},
the ABC model \cite{pleimling}, diffusion with exponential position dependence
of the diffusivity \cite{andrey1}, or dynamics in non-linear maps \cite{julia}.

Here we report a comparative study of Sinai diffusion and ultraslow continuous
time random walks (CTRWs) with super heavy-tailed waiting times \cite{weiss,
denisov,julia,chech}. Previous work focused on ensemble averaged observables.
The routine observation of single particle trajectories in the laboratory forces
us to investigate time averages theoretically. Here we study the time averaged
MSD of Sinai and ultraslow CTRW and demonstrate the fundamental disparity between
ensemble and time averages. This weakly non-ergodic behavior \cite{web,pt} is, up
to a prefactor, identical for both processes. We discuss why, despite the seeming
similarity, the problem of ergodicity in annealed and quenched systems is very
different. In addition, we unveil the \emph{universal\/} exponential fluctuations
of the time averaged MSD of the CTRW process and provide numerical evidence of the
Golosov localization effect for different disorder realizations in Sinai diffusion.
Concurrently, the PDF of the CTRW process is shown to be practically
indistinguishable from that of Sinai diffusion.

\begin{figure}
\includegraphics[width=7.2cm]{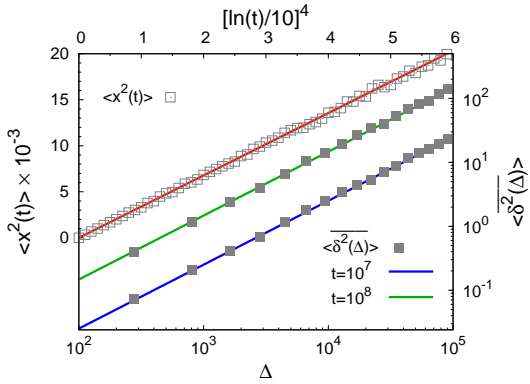}
\caption{Ensemble averaged ($\square$, top and left axes) and time averaged
($\blacksquare$, bottom and right axes, for $t=10^7$ and $t=10^8$) MSD of
Sinai diffusion. The simulations agree excellently with the analytical results
(full lines) of Eqs.~(\ref{sinai_msd}) and (\ref{sinai_tamsd}). Note that we show
$10^{-3}\langle x^2(t)\rangle$ such that for sufficiently long trajectories
[long $t$ in Eq.~(\ref{sinai_tamsd})] $\langle x^2\rangle\gg\overline{\delta^2}$.}
\label{fig1}
\end{figure}

\emph{Sinai diffusion.} We start with the analysis of Sinai diffusion. In the
continuum limit its MSD reads \cite{ising,ledoussal}
\begin{equation}
\label{sinai_msd}
\widetilde{\langle x^2(t)\rangle}\simeq\frac{61}{180}\ln^4(t),
\end{equation}
in scaled units. Here $\langle\cdot\rangle$ is an ensemble average over white
Gaussian noise and $\widetilde{\cdot}$ a disorder average. Fig.~\ref{fig1} shows
excellent agreement with our extensive simulations.

To simulate Sinai diffusion, we follow the approach of Ref.~\cite{ledoussal}
based on a discrete-space Fokker Planck equation for a particular realization of
the random potential. Mapping the Fokker Planck equation onto the corresponding
(imaginary time) Schr{\"o}dinger equation, we obtain the propagator as an
expansion in eigenstates from diagonalization of the Schr{\"o}dinger operator.
From the disorder-averaged one-point and two-point propagators we evaluate
the time and ensemble averaged MSDs (see also below). Results were averaged over
5000 realizations of the random potential with $10^8$ time steps each \cite{long}.

Individual time series $x(t)$ garnered by modern single particle tracking
techniques or from simulations are typically evaluated in terms of the time
averaged MSD \cite{pt}
\begin{equation}
\label{tamsd}
\overline{\delta^2(\Delta)}=\frac{1}{t-\Delta}\int_0^{t-\Delta}\Big[x(t'+\Delta)-
x(t')\Big]^2dt',
\end{equation}
with the lag time $\Delta$ and the length $t$ of the time series. The overline
$\overline{\,\,\cdot\,\,}$ denotes the time average.
Averaging Eq.~(\ref{tamsd}) over many trajectories embedded in different
realizations of the disorder we have
\begin{eqnarray}
\nonumber
\widetilde{\left<\overline{\delta^2(\Delta)}\right>}&\sim&\frac{1}{t-\Delta}
\int_0^{t-\Delta}\left[\langle\widetilde{x^2(t'+\Delta)}\rangle+\langle
\widetilde{x^2(t')}\rangle\right.\\
&&\left.-2\widetilde{\langle x(t')x(t'+\Delta)\rangle}\right]dt'
\label{eatamsd}
\end{eqnarray}
where the two-point position correlation function
\begin{equation}
\widetilde{\langle x(t')x(t'+\Delta)\rangle}=\sqrt{\langle\widetilde{x^2(t'+
\Delta)}\rangle\langle\widetilde{x^2(t')}\rangle}f(y)
\end{equation}
is known from renormalization group calculations by Le Doussal et
al.~\cite{ledoussal}, where $y=\ln(t'+\Delta)/\ln t'$ and
\begin{eqnarray}
\nonumber
f(y)&=&\frac{72}{61y}-\frac{40}{61y^2}-\frac{180}{427y^3}+\frac{2045}{1281y^4}\\
&&+e^{y-1}\left(\frac{20}{61y^2}-\frac{80}{183y^3}-\frac{36}{61y^4}\right).
\end{eqnarray}
Inserting into Eq.~(\ref{eatamsd}) we obtain our first main result, the time
averaged MSD
\begin{equation}
\label{sinai_tamsd}
\widetilde{\left<\overline{\delta^2(\Delta)}\right>}\simeq\frac{3721}{17080}
\ln^4(t)\frac{\Delta}{t}=\langle\widetilde{x^2(t)}\rangle\frac{549}{854}\frac{
\Delta}{t},
\end{equation}
up to corrections of order $(\Delta/t)^2$ \cite{long}. In Eq.~(\ref{sinai_tamsd})
we observe the distinct disparity between the MSD (\ref{sinai_msd}) and its time
averaged analog (\ref{sinai_tamsd}). In contrast to the logarithmic time dependence
(\ref{sinai_msd}) for the MSD, the time averaged MSD (\ref{sinai_tamsd}) grows
\emph{linearly\/} with the lag time $\Delta$. Concurrently, it displays the aging
dependence
proportional to $\ln^4(t)/t$, i.e., the process is progressively slowed down.
Physically, this is effected when the random walker hits increasingly deeper wells.
In Fig.~\ref{fig1} without fit we see excellent agreement between the analytical
prediction (\ref{sinai_tamsd}) and the simulations, including the dependence on
$t$.

\begin{figure}
\includegraphics[width=7.2cm]{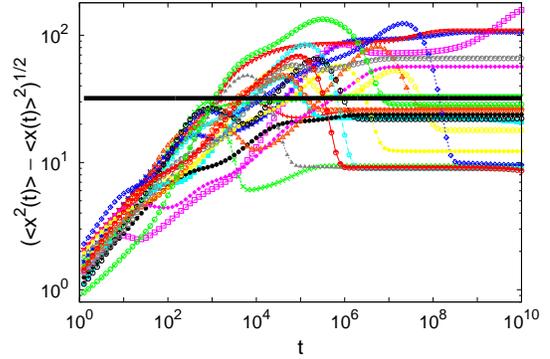}
\caption{Standard deviation $[\langle x^2(t)\rangle-\langle x(t)\rangle^2]^{1/2}$
for individual realizations of the quenched potential landscape in the Sinai model.
The universal localization due to the Golosov effect is distinct. The horizontal
black line shows the value 32.}
\label{fig2}
\end{figure}

However, when discussing time averages in non-trivial quenched systems such as
Sinai landscape, we must distinguish between at least two averaging scenarios.
In the first case each path is realized on its own unique quenched landscape,
and after averaging the individual time averaged MSDs $\overline{\delta^2}$ over
disorder we get Eq.~(\ref{sinai_tamsd}). Second, we can consider one unique
disordered system with an ensemble of non-interacting particles which all start
at the origin. The thermal noise for
each particle is independent. Still, in the long time limit we can expect that
the time averaged MSD for all particles will depend on the specific realization of
the disorder, and due to the Golosov localization effect all trajectories will
yield similar values for the time averaged MSD. Hence we now focus on the
Golosov effect \cite{golosov}: In a given realization of the disorder the
ratio $\langle x(t)\rangle/\ln^2t$ for a fixed initial position up to a prefactor
of order unity becomes deterministic: the particles get stuck in the deepest
potential minimum they can reach within the time $t$. In particular, for the
standard deviation the result $[\langle x^2(t)\rangle-\langle x(t)\rangle^2]^{
1/2}\simeq32$ in scaled units was obtained \cite{golosov}. Fig.~\ref{fig2}
demonstrates that indeed a universal localization of this MSD emerges. The
variations of the onset and height of the plateaus in Fig.~\ref{fig2} reflect
different realizations of the disorder.
An important feature of Sinai diffusion is the slow convergence to the Golosov
localization, found after $10^4$ to $10^5$ time steps. The numerical
analysis of the Golosov effect in different disorder realizations is our
second main result.

\begin{figure}
\includegraphics[width=7.2cm]{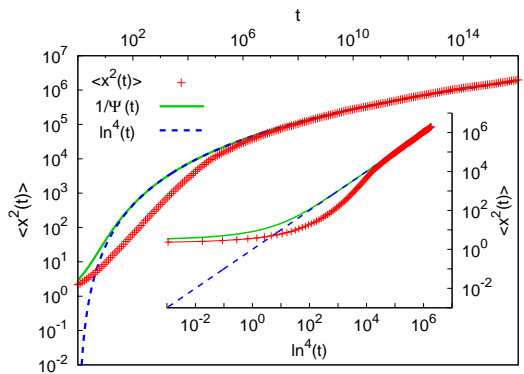}
\caption{MSD $\langle x^2(t)\rangle$ of the CTRW ($\gamma=4$, $\tau_0=e$,
$\langle\delta x^2\rangle=1$ used for all CTRW plots). Main graph:
log-log scales. Inset: $\langle
x^2(t)\rangle$ versus $\ln^4(t)$ with linear asymptote.
\label{ctrws_msd}}
\end{figure}

\emph{Ultraslow CTRW.} In contrast to systems with quenched disorder, those with
annealed disorder typically allow for a rigorous treatment. It would therefore be
desirable to have a process, that captures the essential features of the Sinai
diffusion. We show that such a process is given by a renewal CTRW on a
one-dimensional lattice with the asymptotic form $\psi(\tau)\sim h(\tau)/\tau$ of
the distribution of waiting times $\tau$ elapsing between successive jumps
\cite{denisov}. Here $h(\tau)$ is a slowly varying function \cite{REMMM}. Jumps to
left and right are equally likely, and the probability that no jump occurs until
time $t$ is $\Psi(t)=\int_t^{\infty}\psi(\tau)d\tau$, in terms of which
we express our results. A typical example is the logarithmic form
\cite{denisov,weiss,julia,chech}
\begin{equation}
\label{wtd}
\Psi(t)=\ln^{\gamma}(\tau_0)/\ln^{\gamma}(\tau_0+t),
\end{equation}
where $\tau_0>0$ avoids a divergence at $t=0$. The MSD is given by $\langle x^2(t)
\rangle\sim\langle\delta x^2\rangle/\Psi(t)$ \cite{long}, where $\langle\delta
x^2\rangle$ is the variance of jump lengths \cite{report}. The specific form
(\ref{wtd}) recovers the MSD (\ref{msd}) with diffusivity $K_{\gamma}=\langle
\delta x^2\rangle/[2\ln^{\gamma}(\tau_0)]$ \cite{denisov,weiss,julia,REMM}, i.e.,
for $\gamma=4$, we find a Sinai-like diffusion.
Fig.~\ref{ctrws_msd} demonstrates the convergence of the simulations
to the predicted logarithmic behavior.
The agreement, including the prefactor, after some $10^6$ steps becomes excellent.
In the inset of Fig.~\ref{ctrws_msd} we show the linear asymptotic scaling of
$\langle x^2(t)\rangle$ as function of $\ln^4(t)$.

To obtain the time averaged MSD for this ultraslow CTRW process we note that the
MSD (\ref{msd}) can be rewritten as $\langle x^2(t)\rangle=\langle\delta x^2\rangle
\langle n(t)\rangle$ in terms of the average number of jumps $\langle n(t)\rangle$
of the random walker from $t=0$ up to time $t$, and thus $\langle n(t)
\rangle\sim1/\Psi(t)$. For the calculation of $\langle\overline{\delta^2}\rangle$
we need the correlation function $\langle[x(t'+\Delta)-x(t')]^2\rangle$ [see
Eq.~(\ref{tamsd})]. As
individual jumps are independent random variables with zero mean, $\langle[x(t'+
\Delta)-x(t')]^2\rangle=\langle\delta x^2\rangle[\langle n(t'+\Delta)\rangle-\langle
n(t')\rangle]$. The time averaged MSD then is $\langle\overline{\delta^2}\rangle
\simeq\langle\delta x^2\rangle\Delta/[t\Psi(t)]$. For the form (\ref{wtd}) we have
\begin{equation}
\label{ultraslow_tamsd}
\left<\overline{\delta^2(\Delta)}\right>\sim\langle x^2(t)\rangle\Delta/t:
\end{equation}
Ultraslow CTRWs exhibit weak ergodicity breaking with a linear $\Delta$-dependence,
similar to CTRWs with a power-law form for $\psi(\tau)$ \cite{lubelski,he,pt} and
other processes \cite{sbm,andrey,marcin}.
In particular, Eq.~(\ref{ultraslow_tamsd}), up to the numerical factor $549/854$,
equals the
relation (\ref{sinai_tamsd}) between $\langle x^2\rangle$ and $\overline{\delta^2}$
for the disorder-averaged Sinai diffusion. This similarity is our third main result.

\begin{figure}
\includegraphics[width=7.2cm]{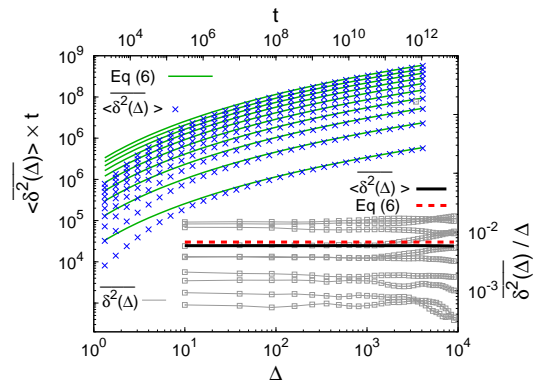}
\caption{Numerical results for $\langle\overline{\delta^2}\rangle\times t$ of the
CTRW
versus $t$ for $\Delta=20$ to $200$ in steps of $20$ ($\times$, bottom to top).
Lines show Eq.~(\ref{ultraslow_tamsd}), without adjustable parameter.
Bottom right: $\overline{\delta^2}/\Delta$ versus lag time $\Delta$ with $t=10^7$
for ten different realizations, showing distinct amplitude scatter. Thick lines
represent Eq.~(\ref{ultraslow_tamsd}) and the average $\langle\overline{\delta^2}
\rangle$ of the plotted realizations.
\label{ctrws_tamsd}}
\end{figure}

In Fig.~\ref{ctrws_tamsd} we plot results from our simulations for the time
averaged MSD, demonstrating the convergence of the numerical results to the
expression (\ref{ultraslow_tamsd}) as function of the length $t$ of the time
series. Fig.~\ref{ctrws_tamsd} also depicts the time averaged MSD as function
of the lag time $\Delta$ for 10 individual realizations along with the result
(\ref{ultraslow_tamsd}) for the mean behavior. A significant amplitude variation
is observed between the individual realizations. This implies that, unlike for
Brownian motion, the experimental observation of a single trajectory producing
$\overline{\delta^2}$ does not provide the full information on neither $\langle x^2
\rangle$ nor $\langle\overline{\delta^2}\rangle$.

Time averages of physical observables such as the MSD (\ref{tamsd}) in weakly
non-ergodic
systems remain random variables even in the limit of long measurement times but
have a well-defined distribution \cite{pt,lubelski,he,bel}. To derive these
fluctuations
for the ultraslow CTRW process in terms of the dimensionless variable $\xi=
\overline{\delta^2}/\langle\overline{\delta^2}\rangle$, we make use of the
relation $\overline{\delta^2(\Delta)}/\langle\overline{\delta^2(\Delta)}\rangle=
n(t)/\langle n(t)\rangle$ \cite{he,johannes} and invoke the probability for the
occurrence of $n$ jumps up to time $t$, $p_n(s)=[1-\psi(\tau)]s^{-1}\exp\{n\ln[\psi
(s)]\}$ in Laplace space \cite{hughes,REM}. For the latter, after Laplace inversion
we get $p_n(t)\sim\Psi(t)\exp\{-n\Psi(t)\}$. Finally, after change of variables,
$\phi(\xi)=p_n(\xi)\times dn/d\xi$, we arrive at the exponential form
for the distribution of time averages
\begin{equation}
\label{scatter}
\lim_{t\to\infty}\phi(\xi)\sim\exp(-\xi).
\end{equation}
Remarkably this result is universal in the sense that it is valid for any generic
ultraslow waiting time distribution $\psi(\tau)\sim h(\tau)/\tau$. In particular,
it is independent of the exponent $\gamma$. The maximum of this distribution is at
$\xi=0$, i.e., many realizations of the ultraslow CTRW process do not perform
any jump. Trajectories with many jumps and thus larger $\xi$ values are
exponentially suppressed. The ergodicity breaking parameter \cite{he} is
$\mathrm{EB}=\lim_{t\to\infty}\langle\xi^2\rangle-1=1$.
The universal fluctuations of ultraslow CTRWs are another central result.
For Sinai diffusion this point remains open, as the simulations methods
do not allow us to obtain sufficiently long and many single trajectories to
analyze their amplitude fluctuations. The results for the Golosov effect
(Fig.~\ref{fig2}) show that the simulations times are beyond our reach.

\begin{figure}
\includegraphics[width=7.2cm]{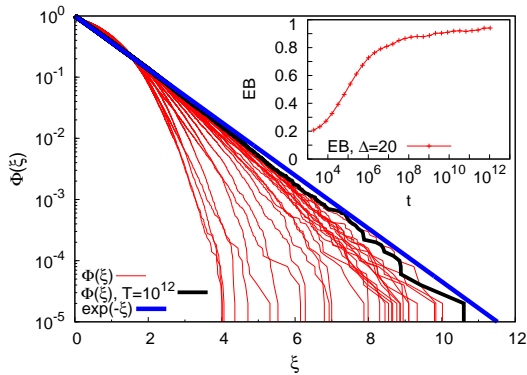}
\caption{The amplitude scatter distribution $\phi(\xi)$ versus $\xi=\overline{\delta
^2}/\langle\overline{\delta^2}\rangle$ for $\Delta=500$ and increasing $t$
successively approaches the exponential (\ref{scatter}) shown by the full line,
compare the realization with $t=10^{12}$. Inset:
convergence $\mathrm{EB}\to1$ of the ergodicity breaking parameter versus $t$
for $\Delta=20$.
\label{ctrw_scatter}}
\end{figure}

\begin{figure}
\includegraphics[width=7.2cm]{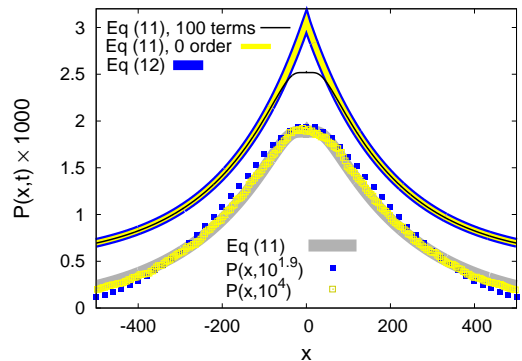}
\caption{PDF (\ref{comtet}) of Sinai diffusion (thick gray line) with numerical
results for two times ($\blacksquare$ and $\square$), showing good convergence.
Shifted curves (top): Rescaled PDF (\ref{ctrw_pdf}) of ultraslow CTRW (thick blue
line) with zero order term of the Sinai PDF (\ref{comtet}) (medium yellow
line) and result (\ref{comtet}) with 100 terms (thin black line),
showing perfect agreement of the tails.
\label{pdf}}
\end{figure}

\emph{Similarity of the PDFs.} We would expect that the PDF $P(x,t)$ is more
sensitive to the deep differences between CTRW and Sinai diffusion.
Fig.~\ref{pdf} shows good convergence of the numerical data to the Sinai PDF
\cite{comtet,nauenberg}
\begin{equation}
\label{comtet}
\widetilde{P(x,t)}\sim\frac{4}{\pi^2\ln^2(t)}\sum_{n=0}^{\infty}\frac{(-1)^n}{2n+1}
\exp\left(-\frac{(2n+1)^2\pi^2|x|}{4\ln^2(t)}\right)
\end{equation}
for increasing $t$. At all times, the zeroth term of this series contributes
$\approx0.33$ to the normalization. Higher order, alternating terms effect a
distinct central plateau, while the wings of the PDF (\ref{comtet}) are
dominated by the zeroth term. The numerical results nicely corroborate the flat
center region of the PDF, which is physically due to the local bias of the Sinai
diffusion at each site effecting the small depletion at the origin. In Fig.~\ref{pdf}
we compare the PDF (\ref{comtet}) with the CTRW PDF
\cite{weiss,denisov,julia}
\begin{equation}
\label{ctrw_pdf}
P(x,t)\sim\sqrt{\Psi(t)/[2\langle\delta x^2\rangle]}\exp\left\{-|x|\sqrt{2\Psi(t)/
\langle\delta x^2\rangle}\right\}.
\end{equation}
Being interested in the qualitative comparison of CTRW and Sinai diffusion in
Eq.~(\ref{comtet}) we rescale the zero order term with a factor of order unity such
that it is identical to the PDF (\ref{ctrw_pdf}) with its distinct cusp at the
origin. Fig.~\ref{pdf} shows that the full result (\ref{comtet}) perfectly matches
the tails of the CTRW PDF (\ref{ctrw_pdf}). In the analysis of experimental data
the PDFs of both processes will be practically indistinguishable unless very
accurate data are available.

\emph{Conclusions.} Sinai diffusion and ultraslow renewal CTRW are two
fundamentally different processes: the former takes place in a quenched random
potential, the latter in an annealed environment. Despite this difference, both
exhibit identical logarithmic scaling of the ensemble MSD and exhibit weak
ergodicity breaking: their time averaged MSD scales linearly with the lag time
$\Delta$
and explicitly depends on the process time $t$ in a characteristic way. We pointed
to the deep connection between the Golosov phenomenon and ergodicity breaking, a
topic that demands rigorous mathematical treatment. The
fluctuations of the amplitude of the time averaged MSD for the ultraslow annealed
model exhibit a remarkable universality: in individual trajectories extended motion
recorded in terms of the time averaged MSD
is exponentially suppressed and details of the waiting time distribution do not
matter. For Sinai diffusion this point remains open as sufficient statistics are
beyond current computational reach.

\acknowledgments

We acknowledge funding from the Academy of Finland (FiDiPro scheme to RM),
Alexander von Humboldt Foundation (AG), and Israel Science Foundation (EB).


\begin{thebibliography}{99}
\bibitem{REMMM} That is, $h(\mu\tau)\sim h(\tau)$ as $\tau\to\infty$ for $\mu>0$.
For instance, $h(\tau)\sim1/\ln^{\gamma}(\tau)$ with $\gamma>0$.
\bibitem{REM} The Laplace transform $f(s)=\mathscr{L}\{f(t);t\to s\}=\int_0^{
\infty}f(t)\exp(-st)dt$ of a function $f(t)$ is denoted by explicit dependence on
the Laplace variable $s$.
\end{thebibliography}
\end{document}